\documentclass[twocolumn,showpacs,preprintnumbers,amsmath,amssymb]{revtex4}
\usepackage{graphicx}
\usepackage{dcolumn}
\usepackage{bm}
\begin{document}

\title{Twisted speckle entities inside wavefront reversal mirrors.}

\author{A.Yu.Okulov}
\email{okulov@kapella.gpi.ru}
\homepage{http://www.gpi.ru/~okulov}
\affiliation{General Physics Institute of Russian Academy of Sciences 
Vavilova str. 38, 119991, Moscow, Russia}

\date{\ June 09, 2009}

\begin{abstract}
The previously unknown property of the optical speckle pattern reported. 
The interference of a speckle with the counter-propagating 
phase-conjugated speckle wave produces 
a randomly distributed ensemble of a twisted entities (ropes) 
surrounding optical vortex lines. These entities appear 
in a wide range of a randomly chosen speckle parameters inside the 
phase-conjugating mirrors regardless to an internal physical 
mechanism of the wavefront reversal. These numerically generated 
interference patterns are 
relevant to the Brillouin PC-mirrors and to a four-wave mixing 
PC-mirrors based upon laser trapped ultracold atomic cloud. 

\end{abstract}


\maketitle

\vspace{1cm}

Phase singularities of the optical beams attracted a substantial interest 
in a recent decades 
from the point of view of optical information processing \cite{Soskin:2001}. 
Helical wavefronts had been shown to affect a processes of the second harmonic generation 
\cite{Soskin:2001}, image processing with a photorefractive 
mirrors \cite{Mamaev:1996}, phase-conjugated reflection  
via nondegenerate four-wave mixing in a cold atoms cloud \cite{Tabosa:1999}. 
The photons with helical wavefunction had been shown to possess 
quantized angular momentum $\hbar$ per photon \cite{Barnett:2002}. 
The optical nonlinearities are capable to transfer the angular momentum 
from photons to an 
ensemble of ultracold atoms \cite{Tabosa:1999}. This effect is considered 
as a possible tool for the light 
storage in addition to the slow light technique based on electromagnetically induced 
transparency (EIT) 
\cite{Fleischhauer:2005,Scully:2001}. Recently the angular momentum transfer 
to BEC cloud of sodium atoms via stimulated Raman scheme had 
been observed \cite{Phillips:2006}. 
Of special interest is the nondegenerate four-wave mixing a 
cold atomic cloud and phase conjugated (PC) reflection 
of the optical vortices (OV), i.e. 
Laguerre-Gaussian beams (LG) with helical wavefronts \cite{Tabosa:1999}. 
The reflection of 
the phase-conjugated LG photons from $10^7$ cesium atoms 
cooled down to $T \approx 10^{-3} K$ and reversal of the orbital angular momentum 
(OAM) had been interpreted as a consequence of internal macroscopic rotations 
inside atomic cloud \cite{Tabosa:2007}. Afterwards 
these experimental results were analyzed from a point of view of 
the angular momentum conservation for the incident and reflected photons from a Brillouin 
PC - mirror \cite{Okulov1:2008}. 
The goal of the current communication is to study the spiral interference patterns 
with the period of $\lambda /2$ around the $nodes$ of optical speckle pattern. 
These coarse interference patterns turn in rotation the Brillouin medium \cite{Okulov1:2008J} 
producing the acoustical vortices carrying OAM \cite{Thomas:2003}. 

The traditional experimental and numerical technique for the visualization 
of the optical phase singularities is based on mixing of a wide quasi-plane wave 
reference optical beam 
with a speckle signal \cite{Zeldovich:1982}. The dark spots 
(zeros of amplitudes) appearing in 
intensity distribution are 
collocated with the helical ramps in phase distribution. 
The current communication reports 
a theoretical description of the alternative 
visualization technique which uses the interference of the speckle wave 
with the counter-propagating 
phase-conjugated one. In fact 
this technique is automatically implemented in a wavefront reversing 
mirrors where a phase-conjugated wave produces near the bright spot 
the Bragg grating 
of dielectric permittivity of the form $cos((k_p+k_s)z)$. 
In the vicinity of the dark line the modulation of the light intensity 
is more complex: the Bragg grating is transformed into spirals 
of the form $cos((k_p+k_s)z \pm 2{\:}\ell \phi)$, 
where $\phi$ is the local azymuthal angle, $\ell$ is topological charge 
of the phase singularity, $k_p , k_s$ are wavenumbers of incident (pump) 
and reflected (Stockes) waves respectively \cite{Okulov1:2008,Okulov1:2008J}. 
In the real PC-mirrors the contrast of such gratings may be 
reduced due to the interference with a 
non-phase-conjugated component of the radiation. 

\begin{figure}
\center{\includegraphics[width=1.7\linewidth , angle=90] {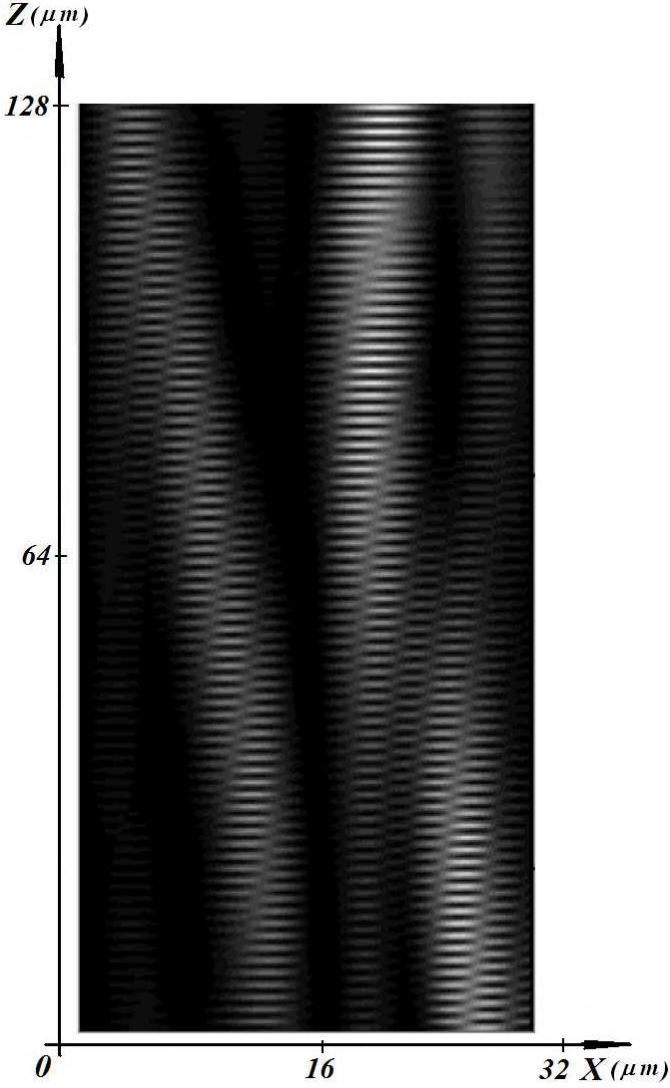}}
\caption{Gray scale intensity plot for a 
small volume of a speckle pattern inside PC-mirror in $(X,Z)$ -plane
($Y=23 \mu m$).
The mean transverse size 
of a speckle is $D \sim 8 \mu m$.
The mean longitudinal length of a speckle entity is 
of the order of the Rayleigh range 
$L_R \approx  D^2/{\lambda}$ for 
$\lambda \sim 1 \mu m$. 
The size of the volume is $32$ $\mu m$ in transverse ($X,Y$) directions 
and $128$ $\mu m$ in longitudinal ($Z$) direction. 
The step of longitudinal modulation $cos((k_{p}+k_{s})z \pm 2\phi)$ induced by 
the interference of 
counter-propagating pump and Stockes waves 
is enlarged here 8 times for visualization purposes. The characteristic 
$\pi$-phase-jump is clearly visible in between adjacent Bragg's $cos$ 
-modulated roll patterns.}
\label{fig.1}
\end{figure}

\begin{figure}
\center{\includegraphics[width=0.9\linewidth] {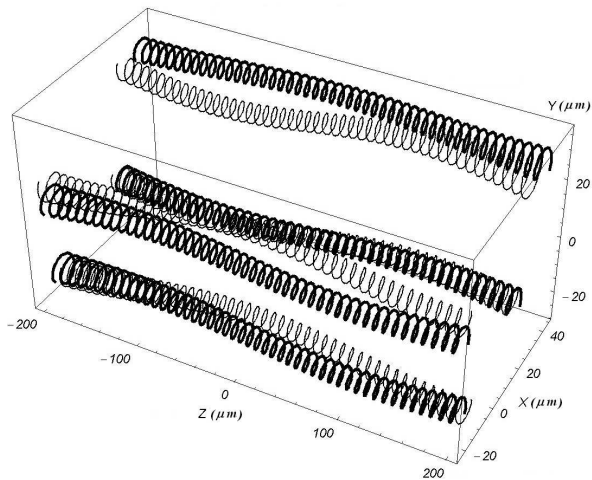}}
\caption{The plot of the light intensity $maxima$ at a given moment $t$ in a 
small volume inside the PC-mirror.
The four pairs of spiral interference patterns with opposite handedness are located 
randomly in space. Their diameters and directions are changed smoothly 
due to diffraction. The step of longitudinal 
modulation $cos((k_{p}+k_{s}) z \pm 2 \phi)$  
is enlarged here 16 times.}
\label{fig.2}
\end{figure}

\begin{figure}
\center{\includegraphics[{width=0.9\linewidth }] {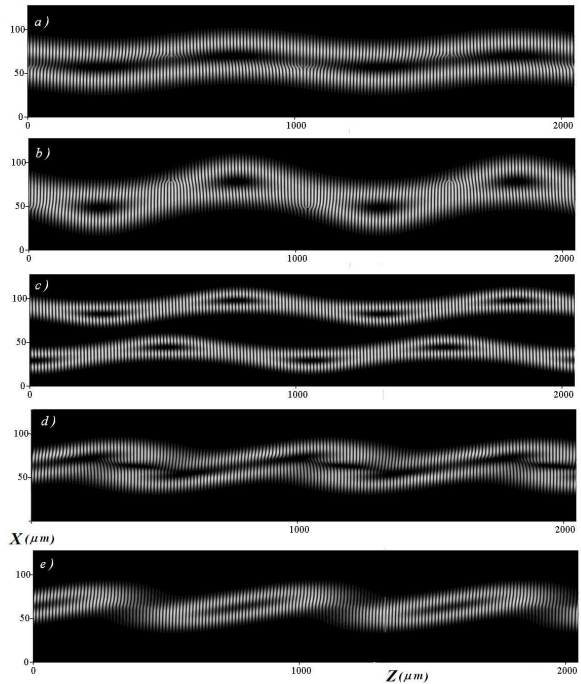}}
\caption
{Gray scale intensity plot for the variational anzatz (\ref{two_vortices}) 
substituted in (\ref{intensity}). The 128 $\times$ 2048 points arrays are used
for the intensity distributions (\ref{intensity}) in 
X $\times$ Z planes at fixed Y=64. The single sinusoidally 
modulated ${LG_0}^1$ vortex line (see eq. (\ref{two_vortices})) 
is shown for $M=0.1 \cdot D$ (a) and $M= D$ (b). 
Two  ${LG_0}^1$ distant 
vortex lines ${LG_0}^1$ (c) with parallel (d) 
opposite (e) topological charges $\pm \ell$ 
constitute the $rope$ when the distance between them is reduced to zero(d,e).
The step of longitudinal 
modulation is enlarged here 32 times.}    
\label{fig.3}
\end{figure}

The novel feature compared to the previous findings \cite{Zeldovich:1982,Zeldovich:1985} 
is that the ideal (or close to an ideal) PC mirror visualizes in the speckle patterns 
a peculiar optical entities, which might be called the $ropes$.
Typically each such a $rope$ is composed of a 2-3 optical vortex lines (fig.\ref{fig.1}). 
In contrast to the well studied straight optical vortex 
lines e.g. LG \cite{Soskin:2001,Barnett:2002}, the OV 
in a speckle field are self-twisted as it seen from fig.\ref{fig.2}. 
Having in mind the well known fact \cite{Zeldovich:1982}, that OV 
appear and annihilate as a pairs of whirls with opposite 
circulations \cite{Okulov1:2008,Okulov1:2008J} it is easy to conclude 
that at least two adjacent OV with opposite topological charges are needed 
to produce a $rope$ (fig.\ref{fig.3}). The mean length of each OV 
in $Z$ direction is the Rayleigh range of a speckle pattern 
$L_R \approx D^2/{\lambda}$, where $D$ 
is an average transverse size (in the plane $X,Y$) of the OV core, 
the wavelength $\lambda=2 \pi /k_p$.
The  fig.\ref{fig.1} shows the numerically generated fragment of a speckle pattern 
which contains three $ropes$ each composed of a set of a vortex lines. 
Let us describe the numerical 
procedure for generating the interference pattern of the two counter-propagating 
speckle fields.

	The standard model of the phase conjugation via stimulated Brillouin scattering (SBS) is the 
Bragg reflection from the sound grating with period $\lambda/2$ moving with 
the speed of sound $v_{ac}$ \cite{Zeldovich:1985}. 
Due to the conservation of momentum $\vec p= \hbar \vec {k_p} = \hbar \vec {k_s} + \hbar \vec {k_{ac}}$ 
the length of the wavevector of sound $k_{ac}$ is equal to doubled length of 
wavevector of pump 
light $k_p$ with an accuracy about $10^{-5}$ \cite{Basov:1980}. The Doppler effect defines the small 
(of the order of $10^{-5}$) frequency shift of the reflected (Stockes) 
wave $\omega_{ac} = 2 \omega_{p} n {v_{ac}}  / c $, where 
$n$ is refractive index of medium, $c$ is the speed of light in vacuum, 
$\omega_{p},\omega_{s}$ is 
the frequency of the pump and Stockes light respectively. 

The evolution in space of the two counter-propagating 
paraxial laser beams inside Brillouin active medium obeys to Maxwell equations 
with the cubic nonlinear polarization induced by electrostrictive effect \cite{Zeldovich:1985}. 
The linearly polarized pump field $E_p$ moves in a positive direction of $Z-axis$,
 the reflected Stockes field $E_s$ with the same polarization propagates in opposite 
direction(fig.1). The acoustic field $Q$ is excited via electrostriction \cite{Zeldovich:1985,Boyd:1990}. 
The three-wave equations for the interaction of $E_p$, $E_s$ with
adiabatic elimination of the $Q \approx E_p {E_s}^{*}$ are:
\begin{equation}
\label{pumpwave}
\ {\frac {\partial {{E_p}(z,x,y,t)}} {\partial z} }
+{\frac {i}{2 k_p}} \Delta_{\bot} {E_p} = -
{\frac { \gamma^2 \omega_p {\:} {k_a}^2} {32 {\rho_0} n c {\:}\omega_{ac}} } |E_s|^2 {E_p}{\:},
\end{equation}
\begin{equation}
\label{stockeswave}
\ {\frac {\partial {{E_s}(z,x,y,t)}} {\partial z} }
-{\frac {i}{2 k_s}} \Delta_{\bot} {E_s} = -
{\frac { \gamma^2 \omega_s {\:} {k_a}^2} {32 {\rho_0} n c {\:}\omega_{ac}} }|E_p|^2 {E_s}{\:},
\end{equation}

where $\gamma=\rho (\partial {\epsilon}/ \partial {\rho})_S$ is the 
electrostrictive coupling constant, 
$ {\rho_0} $ is the unpertubed density of medium \cite{Boyd:1990}.
	Consider the phase conjugation with a random phase plate which 
produces the chaotic transverse modulation of the phase of the incident $E_p$  
with a characteristic size of the transverse inhomogeneity 
of the order of $5-50 \mu m$ \cite{Zeldovich:1985}. In such a case the complex envelope 
amplitude of the inhomogeneous speckle field $E_p$ at the entrance to PC-mirror 
is given as a multimode random field \cite{Okulov:1991}:
\begin{eqnarray}
\label{tetraedral}
\ {E_p}(\vec{r},z=0 ) \approx  { {E_{p}}^o} {\:}{\:} {\sum_{0<{{j_x} ,{j_y}}<N_g}} 
 A_{{j_x} ,{j_y}}{\:}
& &  \nonumber \\
\times
exp{\:}[ {\:}i2 {\pi} \lbrace  {\frac {x {\cdot} {j_x}} {p_x}} {\kappa_{j_x}} 
+{\frac {y \cdot  {j_y}} {p_y} }{\kappa_{j_y}} + i \theta_{{j_x} ,{j_y}} \rbrace ],
\end{eqnarray}
where random phases $\theta_{{j_x} ,{j_y}} $ are the random numbers from interval $[ 0, \pi ]$, 
$ A_{{j_x} ,{j_y}} $ are the real amplitudes of the spatial harmonics, 
$p_x , p_y $  are maximal transverse sizes (in $X, Y $-plane),${\:}$ $\vec{r}=(x, y)=(r, \phi) $, 
${\:}$ ${{j_x} ,{j_y}}$ are integers 
corresponding to $N_g=(16,32,64)$ plane waves with random phases $\theta_{jx,jy}$. 
The random numbers ${\kappa_{j_x}}=p_x/{\tensor{p}{^{'}_{jx}}}$ 
and ${\kappa_{j_y}}=p_y/{\tensor{p}{^{'}_{jy}}}$ having the uniform distribution in the small vicinity 
of a $\bf 1$ are responsible for a 
random tilt of a plane waves constituting the spatial Fourier spectrum 
of the light transmitted through a random phase plate.
For the paraxial beam propagation the amplitude 
and phase structure of the complex field $ E_p$ 
have the following form \cite{Okulov:1990} in an arbitrary $Z$-plane :
\begin{eqnarray}
\label{pump1}
{ E_p}(\vec{r},z>0 ) \approx {{E_{p}}^o}{\:}  {\:}exp{\:}(i k_p z) {\:} \sum_{0<{{j_x} ,{j_y}}<N_g} 
{\:}exp {\:}[{\:}i {\:}\theta_{{j_x} ,{j_y}}]
& &  \nonumber \\
A_{{j_x} ,{j_y}} exp{\:}[ {\:}i 2 {\pi} 
\lbrace 
\frac {x {\cdot} {j_x}} {\tensor{p}{^{'}_{jx}}}+ 
\frac {y {\cdot} {j_y}} {\tensor{p}{^{'}_{jy}}}+ 
\frac {z} {2 k_p} (\frac {{j_x} ^2} {{{\tensor{p}{^{'}_{jx}}}}^2 }+ 
\frac {{j_y}^2} {{\tensor{p}{^{'}_{jy}}}^2 }) \rbrace ],
\end{eqnarray}

The interference pattern $I(\vec{r},z)$ is responsible for the sound grating 
in a Brillouin PC-mirror. For the grating 
produced by a linear superposition of the random plane waves we have: 

\begin{equation}
\label{intensity}
{I}(\vec{r},z>0 ) \approx |{{E_{p}(\vec{r},z)} + E_{s}(\vec{r},z)|^2}{\:}, 
\end{equation}
where the phase-conjugated Stockes wave $E_s$ is given by: 

\begin{eqnarray}
\label{Stockes}
{ E_s}(\vec{r},z>0 ) \approx {{E_{p}}^o} exp{\:}(-i k_s z) {\:} \sum_{0<{{j_x} ,{j_y}}<N_g} 
exp[-i {\:}\theta_{{j_x} ,{j_y}}]
& &  \nonumber \\
A_{{j_x} ,{j_y}}exp[-i 2 {\pi} {\:}
\lbrace 
\frac {x {\cdot} {j_x}} {\tensor{p}{^{'}_{jx}}}+ 
\frac {y {\cdot} {j_y}} {\tensor{p}{^{'}_{jy}}}+ 
\frac {z} {2 k_s} (\frac {{j_x} ^2} {{{{p}{^{'}_{jx}}}}^2 }+ 
\frac {{j_y}^2} {{\tensor{p}{^{'}_{jy}}}^2 }) \rbrace ],
\end{eqnarray}
provided  that ${\:} {\:}  E_{s} \approx {E_{p}}^{*}$ ( noiseless PC-mirror). 
Such an approach presumes the ideally perfect phase-conjugation with ultimate 
reflection $R=1$ or interference of the noninteracting fields 
$E_{s} \sim {E_{p}}^{*}$ when the  
right hand sizes of Eq. (\ref{pumpwave},\ref{stockeswave}) are equal to zero. 

The intensity snapshots given by (\ref{intensity}) were obtained numerically 
by summation of the series (\ref{pump1}) and (\ref{Stockes}) 
on standard $Intel$ platform with $dual-core$ 1.86 Ghz processor 
and $1 {\:}Gb$ memory using standard educational software.
The numerical simulations show that optical vortex lines 
$intertwine$ and form $ropes$. The average length of a $rope$ is of 
the order of Rayleigh range $D^2/{\lambda}$. This may 
be interpreted as a consequence of diffractive 
divergence within the angle $\lambda /D$ (fig.\ref{fig.1}). The
longitudinal modulation by the Bragg factor $cos[(k_p + k_s)z \pm 2 \phi]$ 
is accompanied by an additional bending and twisting at  
characteristic wavelength $\lambda_K$. The latter is of 
the order of a several tenth of optical $\lambda$,
resembling Kelvin modes of the vortex lines \cite{Santos:2009,Pitaevskii:1961}. 
The handedness of helical $ropes$ changes randomly 
from one entity to another: the clockwise and counterclockwise ropes 
appear with the equal probability. 
Despite the statistical nature 
of a speckle pattern (fig.\ref{fig.1}) the $ropes$ are $structurally$ $stable$: 
they appear at a different locations and have a variable sizes but their morphology 
is reproduced from a given statistical realization to another one.

The physical interpretation of these numerically observed patterns 
is given in the fig.\ref{fig.3} using the (\ref{intensity}) with 
the electrical fields $E_p={E_s}^{*}$ in the form of a two overlapping 
elementary ${LG_0}^1$ optical vortices:
\begin{eqnarray}
\label{two_vortices}
\ {E_{p}} \approx  exp{\:}(i k_p z-({x_1}^2+{y_1}^2)+i{\ell {\phi_1}}) \cdot \sqrt {({x_1}^2+{y_1}^2)}
& &  \nonumber \\
+  exp{\:}(i k_p z-({x_2}^2+{y_2}^2) \pm i{\ell {\phi_2}}) \cdot \sqrt {({x_2}^2+{y_2}^2)}
{\:},
\end{eqnarray}
where the location of the vortex cores ${(x_1}D=x+M \cdot cos(2\pi z/{\lambda_K}),
({y_1}D=y+M \cdot sin(2\pi z/{\lambda_K}))$ 
and ${(x_2}D=x+M \cdot cos(2\pi z/{\lambda_K}+\pi /2)),
({y_2}D=y+M \cdot sin(2\pi z/{\lambda_K}+\pi /2))$ 
changes sinusoidally with period $\lambda_K$ along $Z$-axis, $\phi_1$ and $\phi_2$ are the 
local azimuthal angles around vortex cores. Such vortex-vortex 
(fig.\ref{fig.3}d) and vortex-antivortex (fig.\ref{fig.3}e) pair 
interference pattern is used here as a variational anzatz in eq.(\ref{intensity}).
The fig.\ref{fig.3} reproduces qualitatively a some features of 
the numerically generated speckle patterns (fig.\ref{fig.1}). 

In summary the nontrivial topology of the multiply 
connected optical speckle patterns was demonstrated numerically. 
Using the interference with the counter-propagating phase-conjugated 
speckle field we have shown the hidden twisted geometry of the 
multimode wave field 
composed of the randomly tilted plane waves 
\cite{Okulov:1991}. Noteworthy the $ropes$ exist without phase-conjugated 
counterpart: the reflection from PC-mirror makes these twisted entities 
visible due to the characteristic 
$\pi$-phase-jump in between the adjacent Bragg's $sinusoidally$
-modulated roll patterns (fig.\ref{fig.1}). 
Thus it seems probable that a 
$rope$ $entities$ exist in a wide class of a $superpositional$ 
physical fields like an electromagnetic or acoustical ones \cite{Thomas:2003}. The 
other most 
evident examples are the blackbody radiation field in a cavity or a 
monochromatic field in a cavity with a 
rough mirrors. Taking into account the phenomena of a so-called $nonlinear$ 
$superposition$ which take place for collisions of the optical solitons,
vortices \cite{Kivshar:2003} and formation 
of the optical vortex lattices \cite{Okulov:2008}
one might expect the finding of a stable $ropes$ 
in a $nonlinear$ 
fields. The relevant examples are photorefractive PC-mirrors \cite{Mamaev:1996}, 
ultracold matter, e.g. 
phase-conjugating PC-mirrors based on nondegenerate four-wave 
mixing in a trapped gases \cite{Tabosa:1999,Tabosa:2007} 
and the situations of the slow \cite{Fleischhauer:2005,Boyd:2003,Scully:2001} 
and the fast light \cite{Eberly:2007,Okulov:1988} propagation in a resonant medium. 

The reported result extends the set of the possible forms 
of the photon's wavefunction. 
In addition to the well known photon wavefunctions 
in the form of the elementary optical vortex 
lines, e.g. $LG$ beams which possess 
a helical wavefronts \cite{Soskin:2001,Barnett:2002}, the twisted 
entities each composed of the several vortex lines offer a new form 
of the photon's wavefunction \cite{Sipe:1995,Pitaevskii:1982}. 

Acknowledgement. 
This work was supported in part by the Russian Fund of Basic Research Grant 08-02-01229.

\end{document}